\documentclass{aa}
\usepackage{graphicx}
\usepackage{txfonts}
\begin{document}


\title{	Collisional  effects on the formation of the second solar spectrum of the   Sr {\sc ii} $\lambda$4078 line}

          \author{C. \textsc{Deb} and M. \textsc{Derouich}}

    \offprints{M.  \textsc{Derouich}}

  \institute{ Sousse University, LabEM (LR11ES34),   ESSTHS,   Lamine Abbassi street, 4011 H.Sousse, Tunisia  
 }

    \titlerunning{Collisional  effects}
    \authorrunning{\textsc{Deb} \& \textsc{Derouich}}

\date{Accepted 25 September 2014}

\abstract
{ One of the challenging theoretical problems in solar physics  is
the   modeling of polarimetric observations in view of   diagnostics of magnetic fields in the solar atmosphere.  }
{This work aims to  provide  key elements to gain better understanding of   the formation of the second solar spectrum of the Sr II $\lambda$4078 line.    }
{ The atomic states are quantified   by the density matrix elements   expressed  in  the basis of the irreducible tensorial operators.   We perform  accurate computation of the  collisional depolarization and polarization transfer  rates   for all levels involved in the Sr II 4078 \AA\ line.   We solve   the statistical equilibrium equations to calculate the linear polarization degree  where: (a)  collisions are completely ignored, (b) collisions are taken into account in the framework of a simplified atomic model, and (c)   collisions are taken into account in the framework of 5-levels and 5-lines atomic model.
}
{We provide  all collisional
rates needed for Sr II $\lambda$4078 line modeling. Although  Sr II $\lambda$4078 line is the resonance line of Sr II   connecting the ground state $^2S_{1/2}$ to the   excited  state $^2P_{3/2}$, we show that its linear polarization  is sensitive to collisions    with neutral hydrogen  mainly because the  metastable $D$-levels   are vulnerable to collisions.  Thus, a 5-levels and 5-lines atomic     model is needed to study this line. We determine a correction   factor   that one must  apply to the value of the
linear polarization  derived from simplified two-level model.  
 }
{ In a certain  range of hydrogen density,  the effect of isotropic collisions between   Sr II ions  with   hydrogen  atoms 
 is important in determining the polarization of the Sr II $\lambda$4078 line. The use of an atomic model that neglects the metastable level 4d of Sr II 
 can induce   errors of up to 25 \% in the value of the scattering polarization of the solar Sr II $\lambda$4078 line.  }
\keywords
{Scattering -- Collisions processes -- Sun: atmosphere -- Sun: magnetic fields -- Line: formation - Polarization} 
 
\maketitle

\section{Formulation of the problem}
The influences of the Sun on the Earth's climate and environment are the source of new developments in astrophysics
regrouped under the name of space weather.   To better understand these influences one should understand the physics of the Sun and thus  should determine with good accuracy the solar magnetic field.  Thanks to a vigorous technological effort, spectro-polarimetric 
observations with   unprecedented precision and sensitivity are now available to 
the solar physics community (e.g. Stenflo \& Keller   1997; Bianda \& Stenflo 2001; Trujillo Bueno et al. 2001; Gandofer 2000, 2002, 2005 ; Malherbe et al. 2007;  L\'opez-Ariste et al. 2009; Bianda et al. 2011; Mili\'c \& Faurobert 2012). 

Diagnostics   of molecular and atomic lines  formed in the solar atmosphere based on the physics of scattering polarization   are powerful tools to learn about   unresolved magnetic fields  which are very present  in the solar plasma (see for instance Trujillo Bueno et al. \cite{Trujillo_04}; Stenflo \cite{Stenflo_04} and the book of Landi Degl'Innocenti  \& Landolfi 2004).     Reliable interpretation   of  the scattering polarization    often requires  solving   numerically   the coupled set of   equations of the radiative transfer and the statistical equilibrium   of a multilevel  atomic model.

In the solar community,     selected lines from different wavelength   regimes are currently interpreted  in terms of magnetic field and the scientific value of their use  is quantitatively estimated. However one   still must confront many serious problems facing rigorous interpretation of the observed polarization of the solar radiation. For instance, the   Hanle  effect of the magnetic field and  isotropic collisions are mixed in the same observable (the polarization state) which makes the interpretation    of the observed polarization  in terms of magnetic field difficult or imprecise or controversial.

  In the case of  two-levels  atom where the polarization 
of the lower level and collisions are neglected,  one shows that the efficiency of the Hanle effect is controlled by the dimensionless parameter $\Gamma = \frac{2 \pi \;  \mu_{B}  \; B \; g_{J}}{h \; A}= \frac{2 \pi \;  \nu_{L}  g_{J}}{A}$.   $\nu_{L}$ is the Larmor frequency, $h$ is the Planck constant,  $B$ is the magnetic field strength, $\mu_{B}$ is the Bohr magneton,  $g_J$ is the 
 Land\'e factor of the $J$-level under consideration and $ A$ is the  Einstein coefficient for the spontaneous emission of the spectral line having $J$ as upper level.
 $\Gamma=1$ corresponds to $B=B_c$ which is the critical magnetic field strength where the Hanle effect is maximal. For $0.1\;B_c<B<10\;B_c$, one may expect a sizable change of the scattering polarization signal with respect to the unmagnetized reference case.   Numerically, one has
  \begin{eqnarray} 
 B_c ( \textrm{Gauss}) &=&  \frac{1.137 \; 10^{-7} \; A \; (\textrm{s}^{-1})}{g_{J}} 
\end{eqnarray} 
According to L-S coupling, one finds that  $g_{J}=\frac{4}{3}$ for the level $^2P_{3/2}$ of the Sr {\sc ii} ion. The Einstein coefficient    is  $A=1.26   \times 10^{8} \;  s^{-1}$   (NIST  database). Thus, for the linear scattering polarization of the Sr {\sc ii} $\lambda$4078 line, $B_c=10.75 $ Gauss.  Therefore, the scattering polarization  of the   Sr {\sc ii} $\lambda$4078 line is of great   practical interest here 
because the magnetic field in the chromospheric level is around tens of Gauss. Mesearment and interpretation of the polarization of the Sr {\sc ii} $\lambda$4078 line gives an interesting tool to determine magnetic field in the chromosphere of the Sun where the solar magnetism is    not very well known.

In fact,  the Hanle effect in the solar  Sr {\sc ii} $\lambda$4078 line has been studied by  Bianda et al . (1998) (see also Bianda 2003). They obtained a  magnetic
field strength   in the range of 5 to 10 Gauss. They concluded that  the Sr {\sc ii} line gives on average values 30\% lower  than the ones obtained via diagnostics based on the Ca {\sc i} $\lambda$4227 line and they remarked that  this discrepancy cannot be regarded as very significant because of  the uncertainties on the collisional rate determination.  They argued  that  in the higher layers of the solar atmosphere,
where Sr {\sc ii} is formed near the solar limb, the collision rate is
low. Looking to Eqs. (11), (12), and (13)  of Bianda et al .
(1998) giving the magnetic field strength  $B$ and the Hanle rotation angle  according  to the theory adopted by   these authors, it seems that  the obained magnetic field is underestimated because the   effect of elastic collisions (represented by $\gamma_c$ in the theory adopted by Bianda et al . (1998)) is underestimated\footnote{This recalls us the    theoretical  
indications that if collisions are neglected   for molecular lines, the derived magnetic field  is underestimated (Berdyugina \& Fluri   (2004); Bommier et al. (2006)).}.   It is possible that, if the collisions are fully taken into account in the framework of a realistic atmoic model, the Sr {\sc ii} line should give magnetic field values similar to these obtained via diagnostics based on the Ca {\sc i} $\lambda$4227 line.  
 \section{Summary of the theory of the depolarizing collisions}
Isotropic collisional interactions  of the  emitting  Sr {\sc ii} atoms  with nearby  hydrogen  atoms   tend to reestablish thermodynamical equilibrium  inside     Sr {\sc ii} levels, i.e. to  equalize   populations of the Zeeman sublevels and to destroy their coherences, and thus partially destroy the atomic polarization. Consequently, a depolarization of the Sr {\sc ii} line arises from these isotropic collisions.

  Collisions of neutral hydrogen with {\it heavy} atoms like \ion{Sr}{ii}  and   \ion{Sr}{i} are not currently accessible to fully quantum 
chemistry treatments.  To argue this statement,  we cite an example concerning  quantum calculations associated with the $P$-level of  the  \ion{Sr}{i} atom. In fact,    the depolarizing rate of $P$-level of  the  \ion{Sr}{i} was calculated with quantum methods by Faurobert-Scholl et al. (1995) and later by Kerkeni (2002).  Interestingly, Kerkeni (2002)   obtained a value about a
factor 2 smaller than Faurobert-Scholl et al. (1995). This disagreement was not expected since in both cases a quantum approach was used.   We believe that
one possible reason for this discrepancy is the significant complication of ab initio
quantum calculations of the interaction potential between H {\sc i}  and {\it heavy}  atoms
like Sr {\sc i} leading to large error bars in the quantum calculations.     Thus,  for heavy atoms or ions like  \ion{Sr}{ii}  and   \ion{Sr}{i}, one needs semi-classical methods to determine the depolarizing collisional rates. However,  quantum chemistry rates associated with    ions having small size (e.g. \ion{Na}{i},   \ion{Mg}{i}, Ca {\sc ii}, etc)  are important in    validating  approached semi-classical methods.  We used the available  quantum chemistry rates associated with Ca {\sc ii} ions to  validate  the semi-classical method of Derouich et al. (2004). In fact,  Ca {\sc ii} is smaller in size  as compared to Sr {\sc ii} ions, and its quantum-chemistry study is accurate and unproblematic; for Ca {\sc ii} and neutral hydrogen collisions, the quantum interaction potentials are extensively reviewed in the literature.   
 
Collisional  depolarization and  polarization transfer rates    associated to levels of Sr {\sc ii}  are obtained from the  carefully tested  
semi-classical method   of Derouich et al. (2004) which was developed for isotropic collisions between neutral hydrogen atoms and simple ions. In fact, the ionized alkaline earth metals Be {\sc ii}, Mg {\sc ii}, Ca {\sc ii}, Sr {\sc ii}, and
  Ba {\sc ii}   are simple ions because they have
only one valence electron above a filled subshell\footnote{In contrast,
the electronic configuration of a complex ion has one or more valence
electrons above an incomplete (open) subshell (see Derouich et al. 2005). }.  Therefore,  the determination of collisional rates using the approached method developed by 
Derouich et al. (2004)  is possible  with a  percentage of error   lower than 10 \% (see Sect. 6 of Derouich et al. 2004). 
 
Derouich et al. (2004)   applied their  semi-classical   method to obtain depolarization and polarization transfer rates for the upper level 5$p \; $ Ê $^2P$    of the Sr {\sc ii} $\lambda$4078 line. The present work provides new   depolarization and polarization transfer rates associated to the 4$d \; $Ê $^2D$ which are indispensable for a rigorous analysis of the polarization of the Sr {\sc ii} $\lambda$4078 line.

 We describe the  Sr {\sc ii}  atomic  states    by the density matrix elements $ \rho_{q}^{k} (J) $  where 
 $0 \; \le \; k \; \le \;2J$ and $-k \; \le \; q \; \le \;k$.  More details about the physical    meaning    of the tensorial order $k$ and the coherence $q$ could  be found   for instance in Omont (1977), Sahal-Br\'echot (1977) and   Landi Degl'Innocenti  \& Landolfi (2004).  
   
 According to Eq. (1) of Derouich et al. (2004),  the variation of     $ \rho_{q}^{k} (J)$     due to isotropic collisions is:  
\begin{eqnarray} \label{eq_1}
\big[\frac{d \; \rho_q^{k} (\ J)}{dt}\big]_{coll} & = &  
  - \big[\sum_{J' \ne J} \zeta (J \to  J') + D^k(J) \big] \times \rho_q^{k} (J)  \\
&& + \sum_{J' \ne J} 
C^k(J' \to  J)   \times \rho_q^{k} (J')  \nonumber 
\end{eqnarray}
where $C^k(J' \to  J) $ are the polarization transfer rates, $D^k(J)$ are the depolarization rates and $\zeta (J \to  J')$ are the fine structure transfer rates  given by (Eq. 4 of Derouich et al. 2003b):
\begin{eqnarray} \label{eq_1}
\zeta (J \to  J')  & = &   \sqrt{ \frac{2J'+1} {2J+1}} \times
C^0(J \to  J')     
\end{eqnarray}
 Thus\footnote{For a given collision, note   that apart from the multiplicity factor
 $ \frac{2J'+1} {2J+1}$, the $C^k (J,J')$ denoted by Landi Degl'Innocenti \& Landolfi (2004) become the
transfer rates $C^k(J' \to  J)$   as obtained here and in Sahal-Br\'echot (1977) (see also Eq. 3 of Derouich et al. 2003b).},
\begin{eqnarray} \label{eq_1}
\big[\frac{d \; \rho_q^{k} (\ J)}{dt}\big]_{coll} & = &  
  - \big[\sum_{J' \ne J} \sqrt{ \frac{2J'+1} {2J+1}}
C^0(J \to  J')     + D^k(J) \big] \times \rho_q^{k} (J)  \nonumber \\  & + & \sum_{J' \ne J} 
C^k(J' \to  J)   \times \rho_q^{k} (J')  
\end{eqnarray}
Now we  denote  the collisional transfer rates $C^k(J' \to  J)$  by  $C^{(k)}_I(J' \to  J)$ if $J' =J_l <  J$ (inelastic collisions)  and we notice   $C^k(J' \to  J)$  by $C^{(k)}_S(J' \to  J)$ if $J' =J_u>  J$ (super-elastic collisions).
\begin{eqnarray} \label{eq_1}
\big[\frac{d \; \rho_q^{k} (\ J)}{dt}\big]_{coll} & = &  
  - \big[ \sum_{J_l \ne J}  \sqrt{ \frac{2J_l+1} {2J+1}} \; 
C_S^0(J  \to  J_l)   \\  & + &  \sum_{J_u \ne J} 
 \sqrt{ \frac{2J_u+1} {2J+1}} \; C_I^0(J  \to  J_u)    \nonumber + D^k(J) \big] \times \rho_q^{k} (J)  \\
&& \hspace{-1.5cm}  + \sum_{J_l \ne J}  
C_I^k(J_l \to  J)   \times \rho_q^{k} (J_l) + \sum_{J_u \ne J} 
C_S^k(J_u \to  J)   \times \rho_q^{k} (J_u)   \nonumber 
\end{eqnarray}
We denote by $ R^k(J)$ the relaxation rates of rank $k$ given by (Eq. (2) of Derouich 2008):
\begin{eqnarray} \label{eq_1}
R^k(J) & = &  
     \sum_{J_l \ne J}  \sqrt{ \frac{2J_l+1} {2J+1}}
C_S^0(J  \to  J_l) \nonumber  \\ &+&    \sum_{J_u \ne J} 
 \sqrt{ \frac{2J_u+1} {2J+1}} C_I^0(J  \to  J_u)   + D^k(J)    
\end{eqnarray}
Then,
\begin{eqnarray} \label{eq_1}
\big[\frac{d \; \rho_q^{k} (\ J)}{dt}\big]_{coll} & = &  
  -  R^k(J) \times \rho_q^{k} (J)  + \sum_{J_l \ne J} 
C_I^k(J_l \to  J)   \times \rho_q^{k} (J_l) \nonumber  \\ & + & \sum_{J_u \ne J} 
C_S^k(J_u \to  J)   \times \rho_q^{k} (J_u)  
\end{eqnarray}
which   is the same as   the Eq. (3) given in Derouich (2008). Physically, the collisional relaxation  rates $ R^k(J)$  correspond  to the lost of the atomic polarization and, in contrast, the transfer rates $C_I^k(J_l \to  J) $ and $C_S^k(J_u \to  J)$ correspond  to the gain of atomic polarization coming from other levels.

   In the literature, there are different definitions of the collisional rates, and these definitions must be taken
into account when writing the  Eq. (9), in order to calculate the polarization signals correctly. For this reason, Eqs 4-9 are necessary for   a reader  who wants to exploit our    paper. For example,  the depolarization rates as defined in our work, as compared to those defined by Kerkeni et al. (2003), are not equivalent.     Derouich et al. (2004) (Sect. 7)   confronted this problem when they wanted to compare their rates with the collisional rates obtained by Kerkeni et al. (2003). In adiition, there is a small difference between the definition of our collisional rates and the definition of the collisional rates given in Landi DeglÕInnocenti \& Landolfi (2004).  
 
 Note that the observed linear polarization of the Sr {\sc ii} $\lambda$4078 line
is the footprint of only the even orders $k$ inside the Sr {\sc ii} atom, and thus
only depolarization and polarization transfer rates with even $k$
are needed to study such a line.  In addition, since the collisions are isotropic, all collisional rates are $q$-independent.  
 
  \section{Calculation of the collisional rates for the modeling of the Sr {\sc ii} $\lambda$4078  $\AA$ line}
Generally speaking,   studying the polarization of the Sr {\sc ii} $\lambda$4078 line    implies   to solve the radiative transfer equations for polarized radiation in a magnetized atmosphere. To be able to solve numerically these equations one should adopt  a simplified atomic model  as the one presented in Figure 1. 
  Particularly, in the framework of the simplified model, one disregards the effects of the metastable $D$-states in the   Sr {\sc ii} $\lambda$4078 line modeling (see Bianda et al. 1998). 
  
Our intention   is to complete these models   by taking into account  the effect of collisions on the $D$-states. 
To this aim, we consider a  realistic multi-levels and multi-lines atomic model of Figure 2 (hereafter the full model). It contains the 5 levels: $S_{1/2}$, $P_{1/2}$, $P_{3/2}$, $D_{3/2}$ and $D_{5/2}$ and 5 lines:  $\lambda$=10915   $\; \AA$, $\lambda$=10327  $\; \AA$, $\lambda$=10036   $\; \AA$, $\lambda$=4216   $\; \AA$, and $\lambda$=4078   $\; \AA$.  

   We apply our collisional numerical code in order to calculate the collisional scattering matrix after integration of the semi-classical differential coupled equations which are derived from the time dependent Schr\"odinger equation. Once the   scattering matrix is obtained for the $P$- and $D$-states of \ion{Sr}{ii}, we determine the transition probabilities in the tensorial irreducible basis. Afterwards, these propabilities are
integrated over impact parameters and Maxwellian distribution of relative velocities to obtain the depolarization and the transfer
rates. We perform calculations by varying the temperature to obtain the best analytical fit to the collisional rates. The precision on the  Sr {\sc ii}  rates should be similar to the precision obtained on the Ca {\sc ii}  rates (Derouich et al. 2004), i.e. better than 10 \%.  

 Derouich et al. 2004 derived the following expressions of    needed depolarization and polarization transfer rates by isotropic collisions of  $P$-levels of Sr {\sc ii} ions with neutral hydrogen\footnote{ Since the impact approximation is well satisfied for
isotropic collisions between neutral hydrogen atoms and  Sr {\sc ii} atoms in the solar photosphere, the collisional
depolarization rates are simply obtained by multiplying the rates
for a binary collision by the hydrogen density.}:  
   \begin{eqnarray} 
 D^2(3/2)=D^2(P_{3/2}) &=& 5.98 \times 10^{-9} \; n_H \big(\frac{T}{5000}\big)^{0.41}  \\
C_I^0(1/2  \to  3/2)  &=&  C_I^0(P_{1/2} \to P_{3/2})  \nonumber \\ &=& 4.62 \times 10^{-9} \; n_H \big(\frac{T}{5000}\big)^{0.41} \nonumber \\
\end{eqnarray} 
  In  addition in the present work we provide  the following new expressions of collisional  rates associated to the  $D$-levels: 
   \begin{eqnarray} 
D^2(3/2) =D^2(D_{3/2}) &=& 2.28 \times 10^{-9} \; n_H \big(\frac{T}{5000}\big)^{0.42} \nonumber \\
D^2(5/2)=D^2(D_{5/2}) &=& 3.08 \times 10^{-9} \; n_H \big(\frac{T}{5000}\big)^{0.41}  \nonumber  \\
C_I^0(3/2  \to  5/2) &=& C_I^0(D_{3/2} \to D_{5/2})  \nonumber \\ &=& 2.35 \times 10^{-9} \; n_H \big(\frac{T}{5000}\big)^{0.45} \nonumber \\
C_I^2(3/2 \to  5/2) &=& C_I^2(D_{3/2} \to D_{5/2})  \nonumber \\ &=& 1.02 \times 10^{-9} \; n_H \big(\frac{T}{5000}\big)^{0.35} \nonumber  
\end{eqnarray} 
 All rates are given in s$^{-1}$. Only the depolarization rates $D^{k=2}(J)$ and inelastic collisional rates $C^{k}_I(J_l \to J)$
 with $k$=0 and 2  are given as a function of neutral hydrogen number
density  $n_H$ (in cm$^{-3}$) and kinetic temperature $T$ (in Kelvins).  Note that, given the low degree of
radiation anisotropy in the solar atmosphere, one can safely neglect the effect of $D^4(5/2)$ on the linear polarization of the Sr {\sc ii} $\lambda$4078 line.  
However, one can retrieve the values of the
superelastic collisional rates  $C_S^{k}(J_u \to J)$ by applying the detailed balance  
relationship:
 \begin{eqnarray} \label{balance}
 C_S^{k}(J_u \to J_l)  &=& \frac{2J_l+1} {2J_u+1} \;  \textrm{exp}( \frac{E_{J_u}-E_{J_l}}{k_BT}) \; C^{k}_I(J_l \to J_u)
\end{eqnarray} 
with $E_{J_u}$ and  $E_{J_l}$ being the energy of the upper level  $J_u$  and the lower level $J_l$ respectively.  $k_B$ is the Boltzmann
constant. 

It is important to notice that the Eq. (\ref{balance}) is valid only in the case of Maxwell distribution of the velocities of the hydrogen atoms but also the  dimensionless collision strength (generally denoted by $\Omega$) must be symmetrical, i.e.  $\Omega (J_u \to J_l)=\Omega (J_l \to J_u)$.  The  term ``collision strength" was originally suggested by Seaton (1953, 1955) and is now universally used.
The collision strength $\Omega$ contains the information about the collisional transition probability  between two given levels. $\Omega$ is ultimately related to the   scattering matrix. In our collisional method, the interaction potential matrix is hermitian which implies that the scattering S-matrix is unitary
and symmetric  (Derouich et al. 2003a). Consequently    $\Omega$ is symmetrical and the Eq. (\ref{balance}) is well satisfied.

By taking $E_{J_u}$ and $E_{J_l}$ from NIST database,  one finds  at T=6000K:
  \begin{eqnarray}  
C_S^0(P_{3/2} \to P_{1/2}) &=&  0.606 \times  C_I^0(P_{1/2} \to P_{3/2})   \nonumber \\
C_S^0(D_{5/2} \to D_{3/2}) &=& 0.713   \times C_I^0(D_{3/2} \to D_{5/2})  \\
C_S^2(D_{5/2} \to D_{3/2}) &=&  0.713  \times  C_I^2(D_{3/2} \to D_{5/2}) \nonumber 
\end{eqnarray}

  \section{Scattering polarization of the  Sr {\sc ii} $\lambda$4078 line }

   \begin{figure}[h]
\begin{center}
  \includegraphics[width=5 cm]{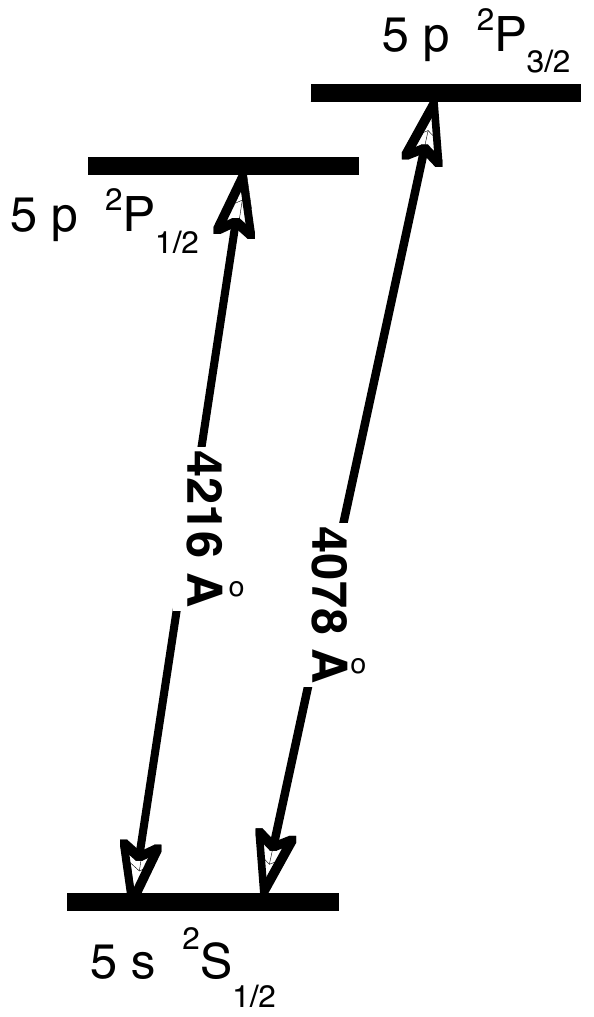}
  \end{center}
\caption{  Partial Grotrian diagram of  Sr {\sc ii} showing the levels and the spectral wavelengths taken into account in the case of the simplified model.
Note that the level spacings are not to scale.}
\label{figure1}
\end{figure}

\begin{figure}[h]
\begin{center}
 \includegraphics[width=7cm]{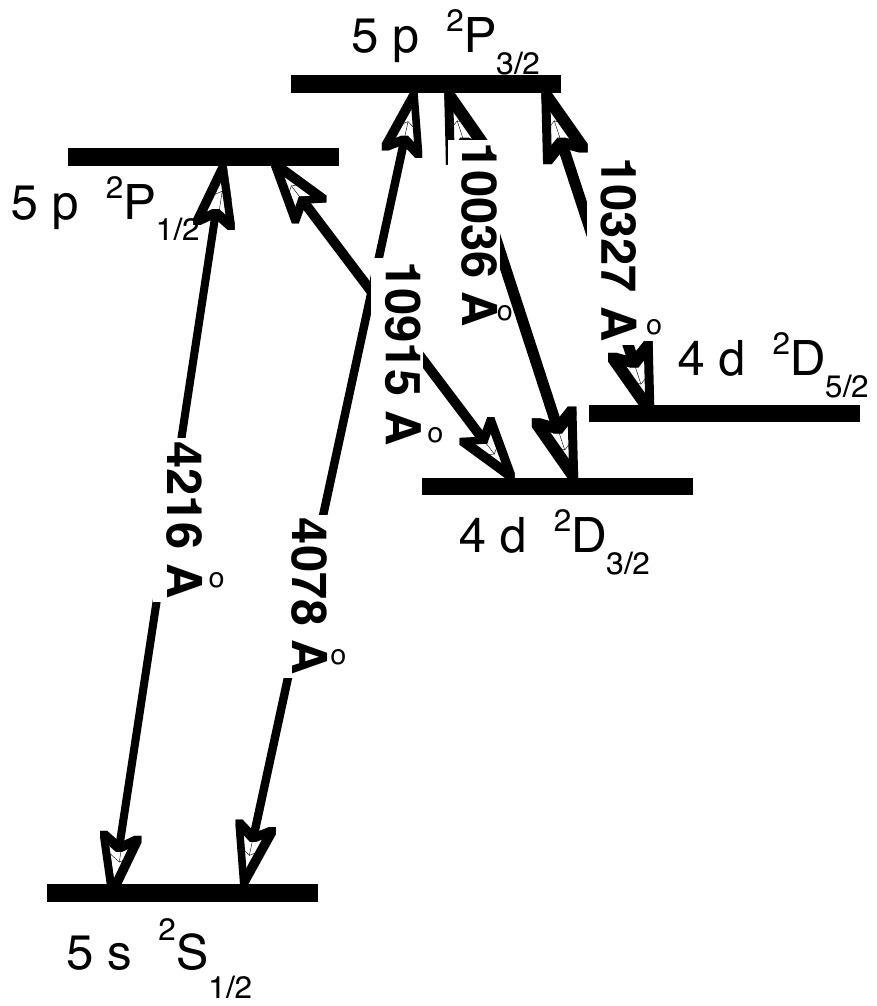}
  \end{center}
\caption{Partial Grotrian diagram of  Sr {\sc ii}  showing the levels and  the allowed radiative transitions used in the full model which takes int account the metastable $D$-states. Note that the level spacings are not to scale.}
\label{figure0}
\end{figure}

  In theory, the core of  the Sr {\sc ii} $\lambda$4078 line is formed  at the chromosphere  but the wings are formed in deeper layers of the photosphere. Since the density of hydrogen atoms $n_H$ in the photosphere is larger than $n_H$ in the   chromosphere, if   the effect of collisions cannot be neglected for the core of  the Sr {\sc ii} $\lambda$4078,  then collisions cannot be neglected  in the
wings of that line.  

\begin{figure}[h]
\begin{center}
\includegraphics[width=8 cm]{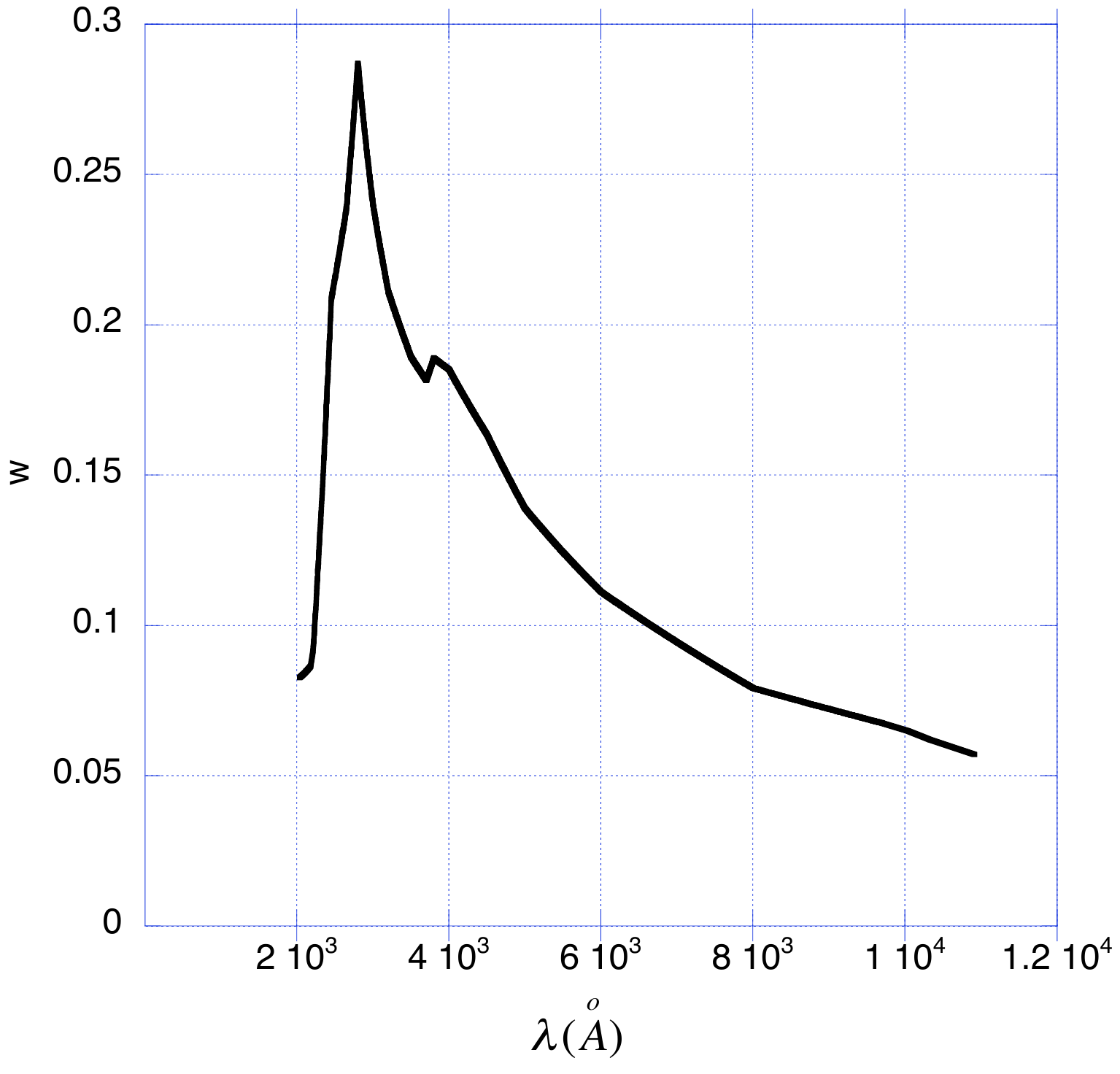}
\end{center}
\caption{ The anisotropy factor  $w(\lambda) = \sqrt{2} \; (J^2_0/J^0_0)$ given as a function of the wavelengths.   }
\label{figure1p}
\end{figure}
\begin{figure}[h]
\begin{center}
\includegraphics[width=8 cm]{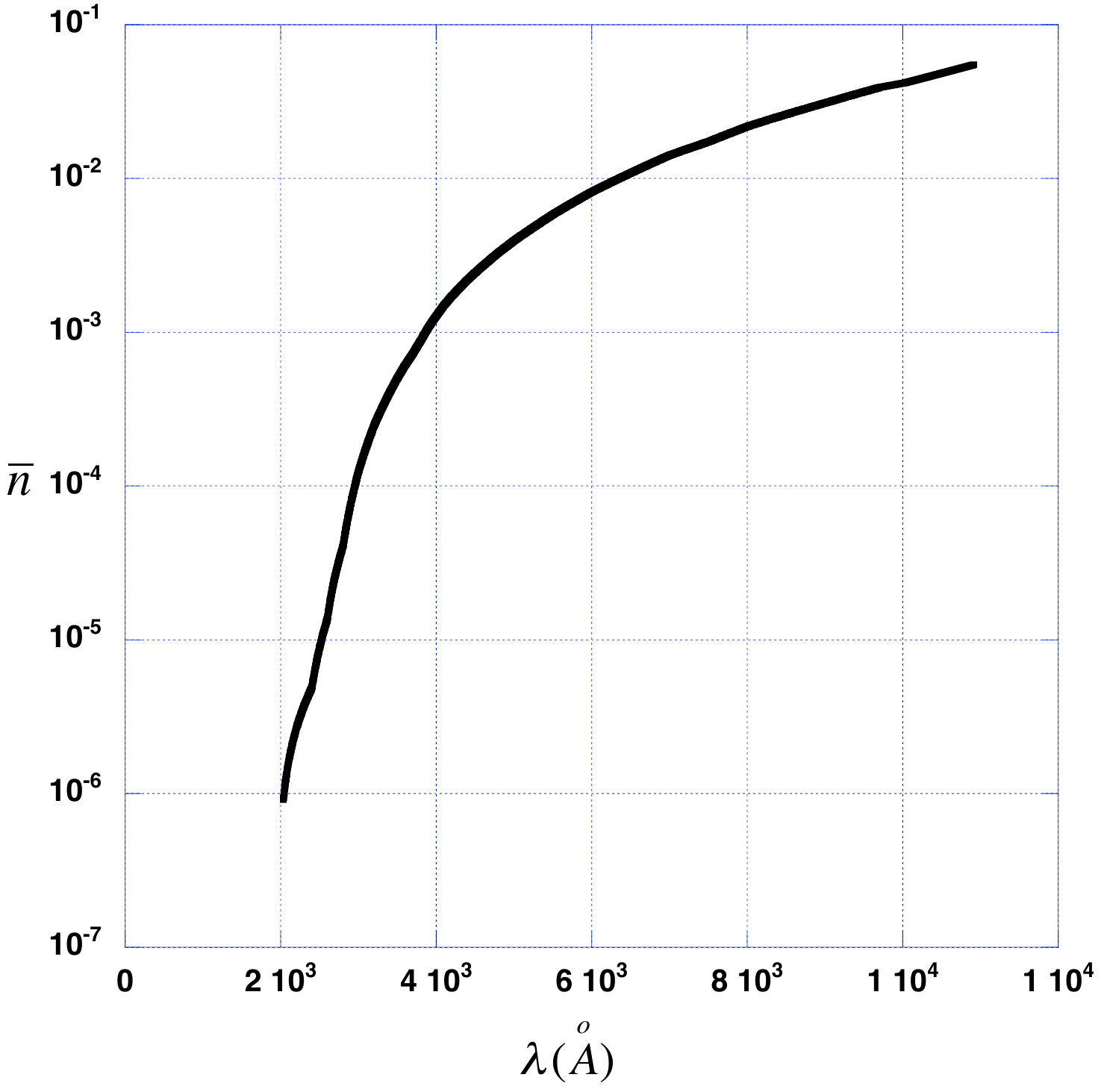}
\end{center}
\caption{ The number of photons per mode  $\bar{n}(\lambda) =J^0_0 \; (c^2/2h\nu^3)$  given as a function of the wavelengths. }
\label{figure1p}
\end{figure}

 We consider that the solar chromosphere is a plane-parallel layer  formed by  Sr {\sc ii} ions which are excited 
anisotropically from below by the photospheric  radiation.  
The components of the incident radiation field at the  wavelenght    $\lambda$=4078\AA\ are usually denoted by  $J_{q}^{k} $  where $k$ is the tensorial order and $q$ represents the coherences in the tensorial basis ($-k \le q \le k$);  the   order  $k$  can be equal to 0  (with $q=0$) or 2 (with $q=0$, $\pm$1, $\pm$2). This radiation field with six components  constitutes a generalization of the  unpolarized   light field where only the quantity $J_{0}^{0}  $ is considered.    In fact, $J_{0}^{0}  $ is proportional to the    intensity of the radiation.  It is useful to notice that if   the chromosphere is assumed to be uniform, the radiation has a cylindrical symmetry around its preferred  direction implying that the coherence components with $q \ne 0$ are zero. In fact, 
$J_{q=\pm1}^{k=2}    $ and $J_{q=\pm2}^{k=2}    $ components quantify the breaking of the cylindrical symmetry around the axis of quantification.

 If the  incident radiation  is no longer anisotropic, the components $J_{q}^{k=2}   $ become  zero which means that   no linear polarization can be created as a result of scattering processes.  
 Regardless    of the  anisotropy of the incident radiation,  the   radiation component associated to the circular polarization usually denoted by $J_{q}^{k=1}$ is negligible. This   means that no odd order $k$ can  be created inside the scattering   strontium ion.  As a result, the Stokes $V$ of the scattered radiation at $\lambda$=4078\AA\ is zero.  

 We concentrate our study on the core of the Sr {\sc ii} $\lambda$4078 line. Neglecting the possible presence of  inhomogeneities of the chromospheric regions, the radiation field is cylindrically symmetrical about the vertical
direction, and its properties are fully described by the two components  $J^0_0(\lambda=4078 \AA)$ and  $J^2_0(\lambda=4078 \AA)$  of the radiation field tensor or, alternatively, by   the value
of the anisotropy factor  $w(\lambda) = \sqrt{2} \; (J^2_0/J^0_0)$   and of the number of photons per mode  $\bar{n}(\lambda) =J^0_0 \; (c^2/2h\nu^3)$ (see e.g. Trujillo Bueno 2001; Manso Sainz \& Landi Degl'Innocenti 2002). $w(\lambda=4078)$ and  $\bar{n}(\lambda=4078)$ are obtained from Figures 3 and 4;    these figures were derived by Manso Sainz \& Landi Degl'Innocenti (2002) from  the center-to-limb variations of the continuum intensities in the quiet Sun given by Cox (2000). 

 \section{The correction factor $f_c$     }
Since we do not solve the radiative transfer equations for the Stokes parameters,   our work should be considered as complementary to other investigations 
which focus on the radiative transfer effects but neglect multilevel and collisional effects. Our aim is to provide   a correction factor on the theoretical value of the  line linear polarization obtained by solving the non-LTE radiative transfer problem in the framework of  two-level modeling. In this sense, we recall that Faurobert et al. (2009) used a correction factor   determined by Derouich (2008) for the Ba {\sc ii} $\lambda$4554 line; they found that, using a correction factor, the accuracy on the magnetic field have been  improved by $\sim$ 50\%.    

The  crorrection  factor $f_c$ must  contain information on  the  difference  between the value of $[p]_{\textrm{simplified}}$ and    $[p]_{\textrm{full}}$. Thus it could be written as:
 \begin{eqnarray} \label{eq_ch3_17}
 f_c & = &   \big( \; [p]_{\textrm{simplified}} - [p]_{\textrm{full}}  \; \big) \times 100
\end{eqnarray}
  Note that  $[p]_{\textrm{simplified}}$  is the linear polarization  inferred from considering the simplified model and  $[p]_{\textrm{full}}$ is the linear polarization  inferred from considering the full model. Obviously,  $[p]_{\textrm{simplified}}$ and $[p]_{\textrm{full}}$  are calculated in the presence of collisions which means that $ f_c$ is a fucntion of  $n_H$.


 We consider that the evolution of the density-matrix elements is due   to the  collisional and radiative effects.    To calculate the linear polarization degree $p=Q/I$ we use the formulae of   Trujillo Bueno (1999):
 \begin{eqnarray}  \label{eq_4}
Q/I && =  \frac{3}{2\sqrt{2}} [\omega^{(2)}_{J_uJ_l} \sigma^{2}_{0} (u) - \omega^{(2)}_{J_lJ_u}  \sigma^{2}_{0} (l)], 
\end{eqnarray}   
which gives the linear polarization degree at the limit of tangential observation, i.e.  at zero altitude above the solar limb. The indices $l$  (for lower) and $u$ (for upper) denote  lower levl $J_l$ and upper $J_u$ of the transition. $\omega^{(k)}_{JJ'}$ is a numerical coefficient  introduced and tabulated for various values of $k$, $J$, and $J'$ in  Landi Degl'Innocenti (\cite{Landi_84}). $\sigma^{k}_{0} = \rho_0^{k} / \rho_0^{0}$, where $ \rho_0^{k}$ are the solutions of the  statistical equilibrium equations (SEE)  including collisional and radiative rates:
 \begin{eqnarray} \label{eq_ch3_17}
\big[\frac{d \; \rho_0^{k} (J)}{dt}\big] & = & \big[\frac{d \; \rho_0^{k} (J)}{dt}\big]_{coll} + \big[\frac{d \; \rho_0^{k} (J)}{dt}\big]_{rad}=0
\end{eqnarray}
    
  For atmospheric level where the hydrogen density $n_H  \sim 10^{15}$ cm$^{-3}$ and $T  \sim 6000 K$, $D^2(P_{3/2})\sim  6.4   \times 10^{6} s^{-1}$ $<<$ $A=1.26   \times 10^{8} \; s^{-1}$.  Because the inverse lifetime of the upper level of the   Sr {\sc ii} $\lambda$4078 line is greater than the value of the elastic depolarizing rate  $D^2(P_{3/2})$, one might think  that the effect of the collisions is negligible. We will demonstrate that at  $n_H  \sim 10^{15}$  the effect 
  of collisions cannot be neglected. 

 We solve the SEE to calculate the $ \rho_0^{k}$ elements and thus we determine the linear polarization $p = Q/I$ of the  Sr {\sc ii} $\lambda$4078 line once by using the simplified atomic model (see Fig. 1) and once by using  the  full model  (see Fig. 2) that accounts for the metastable $D$-levels. We then investigate the sensitivity of the emergent linear polarization to the above-mentioned collisions in order to point  out the range of $n_H$ where the effect of collisions is important.   Since the collisional rates are proportional to the hydrogen density (the impact approximation),
determining the dependence of the polarization to  collisional rates   amounts to the study  of its  dependence on $n_H$.  

Figure 5  represents the linear polarization inferred from the simplified    model in the presence of collisions, $[p]_{\textrm{simplified}}$,  divided  by $[p]_{\textrm{max}}  $   which is  the zero-collisions polarization inferred from the simplified model. In addition,   Fig. 5    shows    the ratio   $   [p]_{\textrm{full}} / [p]_{\textrm{max}} $ giving  the linear polarization inferred from the full model in the presence of collisions   divided  by the same   $[p]_{\textrm{max}}$.  

Looking to the Fig. 5,  the upper horizontal line  with $..   \circ ..$-symbols  represents  the polarization scattered without any sensitive collision effects.  In fact, according to Fig. 5, for $n_H  < 7 \times 10^{14}$ cm$^{-3}$ the variation of $[p]_{\textrm{simplified}}$ is smaller that 2\%, i.e. the ration $   [p]_{\textrm{simplified}} / [p]_{\textrm{max}} \simeq  1 >0.98$. For $n_H  > 7 \times 10^{14}$ cm$^{-3}$,  the ratio  $   [p]_{\textrm{simplified}} / [p]_{\textrm{max}}$ starts to   decrease progressively which reflects the collisional  destruction of the alignement of the $^2P_{3/2}$ level. At $n_H  > 3 \times 10^{17}$ cm$^{-3}$, $   [p]_{\textrm{simplified}} / [p]_{\textrm{max}} <0.1$,    thus the linear atomic polarization in the level $^2P_{3/2}$ is  almost completely destroyed and $   [p]_{\textrm{simplified}} \simeq 0$.

Referring again to the   Fig. 5,  the upper horizontal line  with $- \diamond -$-symbols corresponds to the scattering  polarization obtained using the full model in a range of $n_H$ where the collisions are negligible, i.e. for  $n_H  < 1 \times 10^{14}$ cm$^{-3}$ where $   [p]_{\textrm{full}} / [p]_{\textrm{max}} \simeq  \textrm{constant} >0.93$. It is important to notice that  $   [p]_{\textrm{full}} / [p]_{\textrm{max}} \neq  1$ even in the range where collisions are negligible, this is beacause  the simplified model overestimates the polarization degree leading to the fact that  $   [p]_{\textrm{full}} / [p]_{\textrm{max}} <1$ usually. For $n_H  > 3 \times 10^{14}$ cm$^{-3}$,  the ratio  $   [p]_{\textrm{full}} / [p]_{\textrm{max}}$ starts to   decrease progressively which reflects the collisional  destruction of the alignement of the   $^2D_{3/2}$ and  $^2D_{5/2}$ levels. At $n_H  \simeq  3 \times 10^{14}$ cm$^{-3}$, the alignement of the level $^2P_{3/2}$ is not affected yet by collisions but the polarization of  Sr {\sc ii} $\lambda$4078 line decreases via the destruction of the alignement of the $D$-levels. For $n_H = 4 \times 10^{15}$ cm$^{-3}$, the  $   [p]_{\textrm{full}} / [p]_{\textrm{max}} =0.74$ meaning that the   collisions and multilevels effects  result in a decrease of the polarization degree by more than 25\%; note that  at the same $n_H = 4 \times 10^{15}$ cm$^{-3}$ one has $   [p]_{\textrm{simplified}} / [p]_{\textrm{max}}= 0.9$. 

In order to take into account the  effect of elastic collisions  and of alignment transfer due to multi-level
coupling with the metastable $D$-levels, one must determine the correction factor $f_c$ =   ($ [p]_{\textrm{simplified}}  - [p]_{\textrm{full}}$) $\times 100$ for each $n_H$. For instance at  $n_H = 4 \times 10^{15}$ cm$^{-3}$, $f_c$ =   $(0.9-0.74)  \times 100$=16 \%. Figure 6 gives the variation of $f_c$ as a function of $n_H$. For $\simeq 4 \times 10^{14} <n_H < 5 \times 10^{17}$, the determination of $f_c$  is crucial for a correct determination of the linear polarization. 

\begin{figure}[h]
\begin{center}
\includegraphics[width=8 cm]{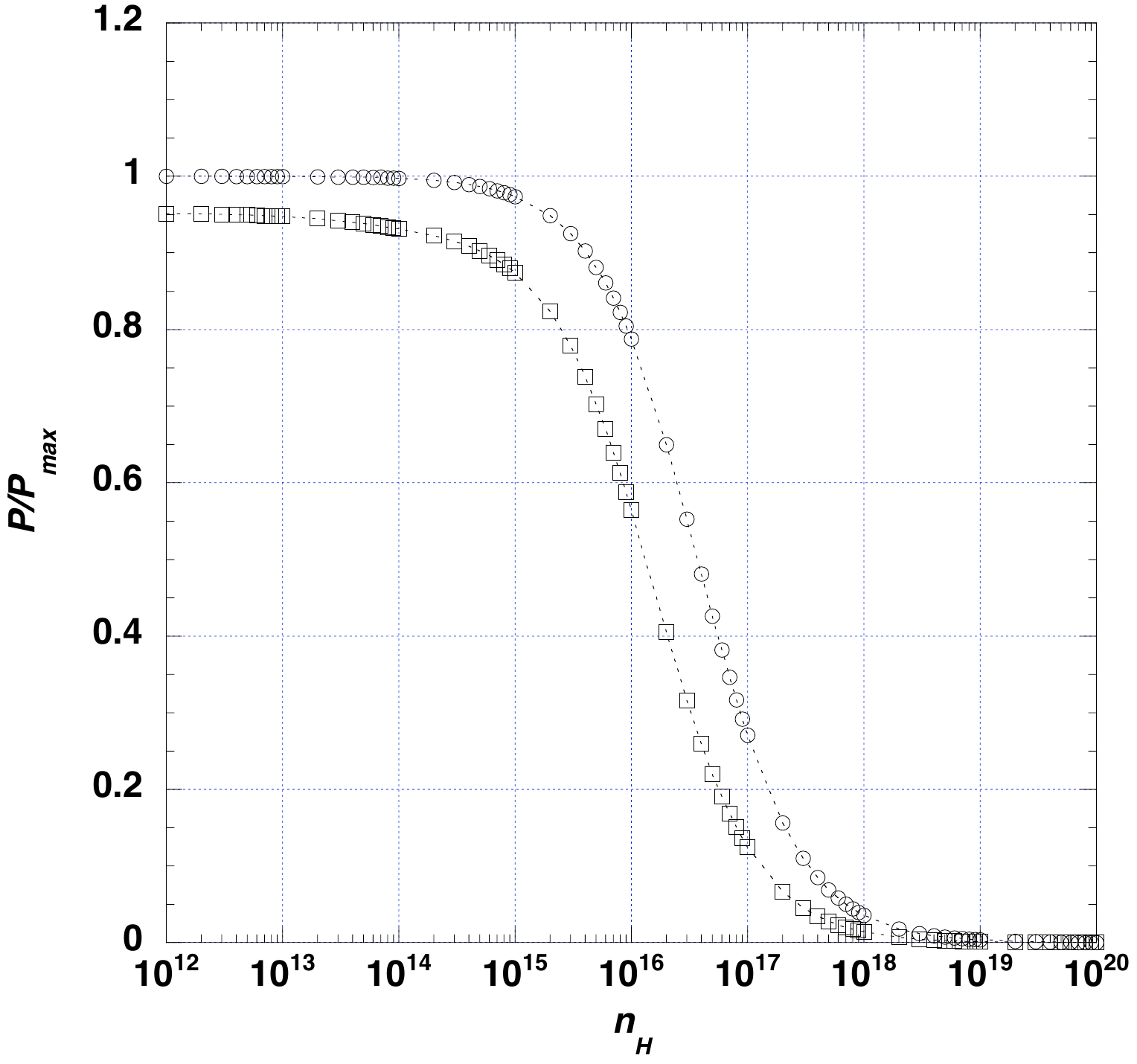}
\end{center}
\caption{Linear polarization ratio $   [p]_{\textrm{simplified}}/ [p]_{\textrm{max}}$ as a function of  $n_H$ plotted as $..   \circ ..$-symbols.   $   [p]_{\textrm{full}}/ [p]_{\textrm{max}}$ as a function of  $n_H$  plotted as  $- \diamond -$-symbols.}
\label{figure1p}
\end{figure}

\begin{figure}
\begin{center}
\includegraphics[width=8 cm]{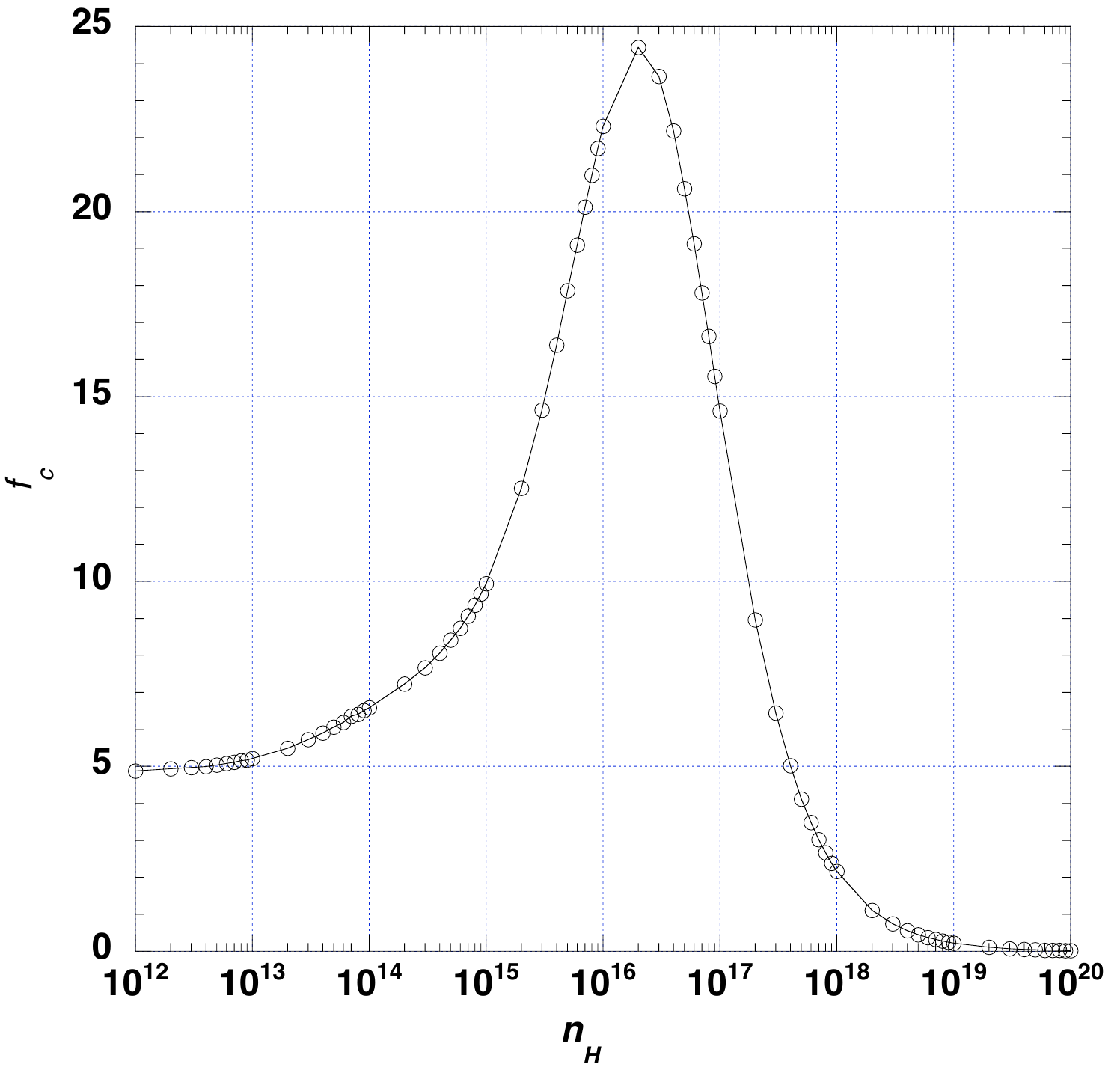}
\end{center}
\caption{Correction factor $f_c$ as a function of $n_H$.  $f_c$    must be applied to the value of the
 linear polarization  derived from the simplified  model.  }
\label{figure1p}
\end{figure}



   \begin{table*}
\begin{center}
\begin{tabular}{cccccc}
\hline
line wavelenghts ($\AA$)     &  10915 &  10327 & 10036     &  4216 &  4078 \\
\hline  
$\bar{n} $     &  $55.028 \times 10^{-3}$ &    $ 46.875 \times 10^{-3}$  & $ 43.115 \times 10^{-3}$    &  $1.757  \times 10^{-3}$ &  $1.450   \times 10^{-3}$\\
 
$w$   & 0.0541  & 0.0636 & 0.06660      & 0.1754  &   0.1816  \\
Einstein Coefficient $A$ (s$^{-1}$) &  $9.5 \times 10^{6}$  & $8.7 \times 10^{6}$ & $1. \times 10^{6}$      &  $1.41  \times 10^{8}$  &      $1.26   \times 10^{8}$ 
\end{tabular}
\end{center} 
\caption{$w$ and $\bar{n}$ are the   anisotropy factor   and the   number of photons per mode for the \ion{Sr}{ii} lines.}
\label{CaII-ModelC}
\end{table*}








 \section{Conclusion}
     We use our collisional approach and our numerical code to calculate new collisional rates associated with the D-states of Sr {\sc ii}, and  using another numerical code, we calculate the polarization of the Sr {\sc ii} line by including  our collisional rates in the statistical equilibrium of a Sr {\sc ii} model atom.    Our aim was to pay particular attention to a subtle collisional depolarizing effect of the Sr {\sc ii} $\lambda$4078 line due to the metastable $D$-levels, this should help   to provide  key elements to analyze quantitatively   the scattering polarization   of the Sr {\sc ii} $\lambda$4078 line. In fact,     we find that  the Sr {\sc ii} $\lambda$4078 line  could be collisionally depolarized even when its upper level   $^2P_{3/2}$ is not affected by collisions.  A  given model can correctly reproduce the observed  scattering polarization  only if one  takes properly into account    the   collisional  perturbation of the atmoic  levels   by the interaction  with nearby hydrogen atoms.     It is important to notice that our results have to be considered as information complementary to the models taking radiative transfer into account but without taking properly into account the effect of collisions in  realistic  multilevel schemes. 
 
As the  scattering polarization degree is usually rather 
small   (e.g. polarization of the order of 1\% in Fraunhofer lines 
in the photosphere and chromosphere of the Sun), careful and rigorous  modelling of that polarization is of fundamental importance for learning especially about a weak and even unresolved 
magnetic field by its Hanle effect.

\end{document}